\documentclass[12pt]{article}
\begin{document}

\title{\bf Matter Inheritance Symmetries of Spherically
Symmetric Static Spacetimes}

\author{M. Sharif \thanks{e-mail: msharif@math.pu.edu.pk}
\\ Department of Mathematics, University of the Punjab,\\
Quaid-e-Azam Campus Lahore-54590, PAKISTAN.}

\date{}

\maketitle

\begin{abstract}
In this paper we discuss matter inheritance collineations by
giving a complete classification of spherically symmetric static
spacetimes by their matter inheritance symmetries. It is shown
that when the energy-momentum tensor is degenerate, most of the
cases yield infinite dimensional matter inheriting symmetries. It
is worth mentioning here that two cases provide finite dimensional
matter inheriting vectors even for the degenerate case. The
non-degenerate case provides finite dimensional matter inheriting
symmetries. We obtain different constraints on the energy-momentum
tensor in each case. It is interesting to note that if the
inheriting factor vanishes, matter inheriting collineations reduce
to be matter collineations already available in the literature.
This idea of matter inheritance collineations turn out to be the
same as homotheties and conformal Killing vectors are for the
metric tensor.
\end{abstract}

{\bf Keywords }: Matter Inheritance Symmetries, Spherically
Symmetric Static Spacetimes.
\date{}

\newpage

\section{Introduction}

Let $(M,g)$ be a spacetime, where $M$ is a smooth, connected,
Hausdorff four-dimensional manifold and $g$ is smooth Lorentzian
metric of signature (+ - - -) defined on $M$. The manifold $M$ and
the metric $g$ are assumed smooth ($C^{\infty}$). We shall use the
usual component notation in local charts, and a covariant
derivative with respect to the symmetric connection $\Gamma$
associated with the metric $g$ will be denoted by a semicolon and
a partial derivative by a comma. A smooth vector field ${\bf \xi}$
is said to generate a Matter inheritance collineation (MIC) [1-2]
if it satisfies the following equation
\begin{equation}
\pounds_{\xi}T_{ab}=\psi(x^a)T_{ab},
\end{equation}
where $\pounds$ is the Lie derivative operator, $\xi^a$ is the
symmetry or collineation vector. If $\psi=0$, then
$\pounds_{\xi}T_{ab}=0$ and $\xi$ is said to preserve a matter
symmetry [3] on $M$ or is said to generate a matter collineation
(MC). If $\xi$ is a Killing vector (KV), then
$\pounds_{\xi}T_{ab}=0$ and hence any isometry is also a MC. The
converse is not true, in general. The MIC Eq.(1) can be written in
component form as
\begin{equation}
T_{ab,c} \xi^c + T_{ac} \xi^c_{,b} + T_{cb} \xi^c_{,a}=\psi
T_{ab}, \quad(a,b,c=0,1,2,3).
\end{equation}
A recent literature [3-13] shows a significant interest in the
study of the various symmetries (in particular, Ricci and matter
collineations). These symmetries arise in the exact solutions of
Einstein's field equations
\begin{equation}
R_{ab}-\frac{1}{2}Rg_{ab}\equiv G_{ab}=\kappa T_{ab},
\end{equation}
where $\kappa$ is the gravitational constant, $G_{ab}$ is the
Einstein tensor, $R_{ab}$ is the Ricci and $T_{ab}$ is the matter
(energy-momentum) tensor. Also, $R = g^{ab} R_{ab}$ is the Ricci
scalar. We have assumed here that the cosmological constant
$\Lambda=0$.

There exists a large body of literature on classification of
spacetimes according to their isometries and the groups admitted
by them [14-17]. These investigations of symmetries played an
important role in the classification of spacetimes, giving rise to
many interesting results with useful applications. As curvature
and Ricci tensors play a significant role in understanding the
geometric structure of metrics, the energy-momentum tensor enables
us to understand the physical structure of spacetimes.

Many of the difficulties encountered in studying such symmetries
in general relativity are due to the fact that the associated
vector fields may constitute infinite dimensional vector space.
Problems of this nature are encountered very rear in the study of
other symmetries. Unlike in the usual study of affine, conformal
and projective symmetry where the initial assumption that $\xi$ is
$C^3$ ($C^2$ for affines) is sufficient to ensure that $\xi$ is
smooth ($C^\infty$), the solution of (2) may be smooth, or $C^m$
(but not $C^{m+1}$) for any $M$ with $1\leq m<\infty$. Hence the
set of solutions of (2), whilst necessarily a vector space, is not
necessarily a Lie algebra under the usual Lie bracket operation.
Of course, if $\xi_{(1)}$ and $\xi_{(2)}$ satisfy (2) then so does
$[\xi_{(1)},\xi_{(2)}]$ provided it is $C^1$. The set of all
smooth MCs on $M$ is, of course, a Lie algebra.

In a recent paper [11], we have classified spherically symmetric
spacetimes according to their MCs. A complete classification of
static plane symmetric and cylindrically symmetric static
spacetimes according to their MCs has been given in very recent
papers [12] and [13] respectively. Some interesting consequences
turn up. Moreover, Duggal [1-2] has proposed a relationship of the
Ricci inheritace symmetry with conformal KV. This paper is devoted
to classify spherically symmetric static spacetimes according to
their MI symmetry. The rest of the paper is organized as follows.
In the next section, we shall write down MIC equations. In section
3, we shall solve these equations for different possibilities to
obtain MIC. Finally, we shall summarize the results in section 4.

\section{Matter Inheritance Equations}

The most general spherically symmetric static metric is given as
\begin{equation}
ds^2 = e^{\nu(r)}dt^2-e^{\lambda(r)}d r^2-e^{\mu(r)}d\Omega^2,
\end{equation}
The non-vanishing components of the energy-momentum tensor
$T_{ab}$ are
\begin{eqnarray}
T_0&=&\frac{e^{\nu-\lambda}}{4}(2\mu'\lambda'-3\mu'^2-4\mu'')
+e^{\nu-\mu},\nonumber \\
T_1&=&\frac{\mu'^2}{4}+\frac{\mu'\nu'}{2}-e^{\lambda-\mu},\nonumber \\
T_2&=&\frac{e^{\mu-\lambda}}{4}(2\mu''+\mu'^2-\mu'\lambda'
+\mu'\nu'+2\nu''+\nu'^2-\nu'\lambda'),\nonumber \\
T_3&=&T_2\sin^2\theta.
\end{eqnarray}
We can write MIC Eqs.(2) in the expanded form as follows
\begin{equation}
T_{0,1} \xi^1 + 2 T_0 \xi^0_{,0} = \psi T_0,
\end{equation}
\begin{equation}
T_0 \xi^0_{,1} + T_1 \xi^1_{,0} = 0,
\end{equation}
\begin{equation}
T_0 \xi^0_{,2} + T_2 \xi^2_{,0} = 0,
\end{equation}
\begin{equation}
T_0 \xi^0_{,3} + \sin^2 \theta T_2 \xi^3_{,0} = 0,
\end{equation}
\begin{equation}
T_{1,1} \xi^1 + 2 T_1 \xi^1_{,1} = \psi T_1,
\end{equation}
\begin{equation}
T_1 \xi^1_{,2} + T_2 \xi^2_{,1} = 0,
\end{equation}
\begin{equation}
T_1 \xi^1_{,3} +\sin^2 \theta T_2 \xi^3_{,1} =0,
\end{equation}
\begin{equation}
T_{2,1} \xi^1 + 2 T_2 \xi^2_{,2} =\psi T_2,
\end{equation}
\begin{equation}
T_2(\xi^2_{,3} + \sin^2 \theta \xi^3_{,2}) =0,
\end{equation}
\begin{equation}
T_{2,1}\xi^1+2T_2\cot\theta\xi^2+2T_2\xi^3_{,3}=\psi T_2.
\end{equation}
For the sake of simplicity, we are using the notation
$T_{aa}=T_a$.

\section{Matter Inheritance Symmetries}

In this section, we solve the MIC equations given by Eq.(6)-(15)
for different situations to find MIC.

From Eq.(14), we have the following two possibilities:
\begin{eqnarray*}
(1)\quad T_2=0,\quad \xi^2_{,3}+\xi^3_{,2}\sin^2\theta\neq0,\nonumber\\
(2)\quad T_2\neq0,\quad \xi^2_{,3}+\xi^3_{,2}\sin^2\theta=0.
\end{eqnarray*}
It is to be noted that the case (1) corresponds to degenerate case
while in the case (2), some of the subcases will lead to
non-degenerate case.\\
\par\noindent
{\bf Case (1):} When we use constraint of this case in MIC
equations, we get
\begin{eqnarray}
T_0\xi^0_{,2}=0,\quad T_0\xi^0_{,3}=0,\quad T_1\xi^1_{,2}=0,\quad
T_1\xi^1_{,3}=0.
\end{eqnarray}
The last two equations give rise to the following two subcases:
\begin{eqnarray*}
(a_1)\quad T_1=0,\quad \xi^1_{,2}\neq0\neq\xi^1_{,3},\\
(a_2)\quad T_1\neq0,\quad \xi^1_{,2}=0=\xi^1_{,3}.
\end{eqnarray*}
{\bf Subcase 1$(a_1)$:} The subcase 1$(a_1)$ yields further two
possibilities:
\begin{eqnarray*}
(b_1)\quad T_0=0,\quad \xi^0_{,2}\neq0\neq\xi^0_{,3}\neq\xi^0_{,1},\\
(b_2)\quad T_0\neq0,\quad \xi^0_{,2}=0=\xi^0_{,3}=\xi^0_{,1}.
\end{eqnarray*}
The first possibility 1$a_1(b_1)$ gives arbitrary MI vectors while
the second possibility 1$a_1(b_2)$ leads to a contradiction.\\
\par\noindent
{\bf Subcase 1$(a_2)$:} When we solve MIC equations for the
subcase 1$(a_2)$, we obtain the following two options
\begin{eqnarray*}
(b_1)\quad T_0=0,\quad \xi^0_{,2}\neq0\neq\xi^0_{,3},\\
(b_2)\quad T_0\neq0,\quad \xi^0_{,2}=0=\xi^0_{,3}.
\end{eqnarray*}
In the first case 1$a_2(b_1)$, i.e., $T_0=0,~
\xi^0_{,2}\neq0\neq\xi^0_{,3}$, MIC equations yield the following
solution
\begin{eqnarray}
\xi^l&=&\xi^l(x^a),\quad l=0,2,3,\nonumber\\
\xi^1&=&\frac{1}{\sqrt{T_1}}(\frac{\psi}{2}\int{\sqrt{T_1}dr+k}),
\end{eqnarray}
where $k$ is an integration constant.

For the option 1$a_2(b_2)$, simultaneous solution of MIC equations
gives the following constraint
\begin{eqnarray}
\frac{B_{,000}}{\dot{B}}=k=\frac{T_0}{\sqrt{T_1}}
(\frac{T'_0}{2T_0\sqrt{T_1}})'
\end{eqnarray}
where $B$ is an arbitrary integration function of $t$ and $k$ is a
separation constant. $B_{,0}$ and $\dot{B}$ are the same notations
and represent derivative with respect to time. This gives us three
more possibilities according to
\begin{eqnarray*}
(c_1)\quad k>0,\quad (c_2)\quad k=0,\quad (c_3)\quad k<0.
\end{eqnarray*}
In the first case 1$a_2b_2(c_1)$, we obtain the following solution
\begin{eqnarray}
\xi^0&=&\frac{\psi}{2}t-\frac{\psi T'_0}{4T_0\sqrt{T_1}}
\int{\sqrt{T_1}dr}\nonumber\\
&-&\frac{T'_0}{2T_0k\sqrt{T_1}}
[kc_0t+c_1\cosh\sqrt{k}t-c_2\sinh\sqrt{k}t]+c_3,\nonumber\\
\xi^1&=&\frac{\psi}{2\sqrt{T_1}}\int{\sqrt{T_1}dr}\nonumber\\
&+&\frac{1}{\sqrt{T_1}}
[c_0+c_1\cosh\sqrt{k}t+c_2\sinh\sqrt{k}t]+c_3,\nonumber\\
\xi^2&=&\xi^2(x^a),\quad \xi^3=\xi(x^a),
\end{eqnarray}
where
\begin{eqnarray*}
(\frac{T'_0}{4T_0\sqrt{T_1}})'=-k\frac{\sqrt{T_1}}{T_0},\quad
(\frac{\psi T'_0}{4T_0\sqrt{T_1}}\int{\sqrt{T_1}dr})'
=-c_0k\frac{\sqrt{T_1}}{T_0}.
\end{eqnarray*}
For the case 1$a_2b_2(c_2)$, Eq.(18) yields
\begin{eqnarray}
B(t)=k_1\frac{t^2}{2}+k_2t+k_3,\quad
\frac{T'_0}{2T_0\sqrt{T_1}}=m,
\end{eqnarray}
where $m$ is an arbitrary constant which gives further three
options according to
\begin{eqnarray*}
(d_1)\quad m>0,\quad (d_2)\quad m=0,\quad (d_3)\quad m<0.
\end{eqnarray*}
The possibilities 1$a_2b_2c_2(d_1)$ and 1$a_2b_2c_2(d_3)$ lead to
contradiction. The possibility 1$a_2b_2c_2(d_2)$ gives the
following MI vectors
\begin{eqnarray}
\xi^0&=&\frac{\psi}{2}t+k_4,\nonumber\\
\xi^1&=&\frac{\psi}{2}t-\int{\sqrt{T_1}dr}+\frac{k_3}{\sqrt{T_1}},\nonumber\\
\xi^j&=&\xi^j(x^a),\quad j=2,3.
\end{eqnarray}
The case 1$a_2b_2(c_3)$ gives the similar results as the case
1$a_2b_1(c_1)$. We see that all the possibilities of the case 1
yield infinite dimensional MIC as it is expected for the
degenerate energy-momentum tensor.\\
\par\noindent
{\bf Case 2:} Solving MIC Eqs.(11), (12), (16) and (17)
simultaneously, we obtain
\begin{equation}
T_1(\xi^1_{,23}-\cot\theta\xi^1_{,3})=0,
\end{equation}
From here, it follows that either
\begin{eqnarray*}
(a_1)\quad T_1=0,\quad \xi^1_{,23}-\xi^1_{,3}\cot\theta\neq0,\nonumber\\
or\quad(a_2)\quad T_1\neq0,\quad
\xi^1_{,23}-\xi^1_{,3}\cot\theta=0.
\end{eqnarray*}
{\bf Subcase 2$(a_1)$}: In this case, it is obvious from MIC
Eq.(7) that we have the following two possibilities:
\begin{eqnarray*}
(b_1)\quad T_0=0,\quad \xi^0_{,1}\neq0,\\
(b_2)\quad T_0\neq0,\quad \xi^0_{,1}=0.
\end{eqnarray*}
When we consider the case 2$a_1(b_1)$, we obtain the following
solution
\begin{eqnarray}
\xi^0&=&\xi^0(x^a),\nonumber\\
\xi^i&=&\xi^i(\theta,\phi),\quad i=1,2,3.
\end{eqnarray}
For the case 2$a_1(b_2)$, when we solve MIC equations, we arrive
at the following result
\begin{eqnarray*}
(\frac{T_0}{T_2})'\xi^0_{,2}=0
\end{eqnarray*}
which yields the following two options
\begin{eqnarray*}
(c_1)\quad (\frac{T_0}{T_2})'=0,\quad \xi^0_{,2}\neq0,\\
(c_2)\quad (\frac{T_0}{T_2})'\neq0,\quad \xi^0_{,2}=0.
\end{eqnarray*}
In the case 2$a_1b_2(c_1)$, solution of MIC equations gives the
following MI vectors
\begin{eqnarray}
\xi_{(0)}&=&\partial_t,\nonumber\\
\xi_{(1)}&=&\sin\phi\partial_\theta+\cot\theta\cos\phi\partial_\phi,\nonumber\\
\xi_{(2)}&=&\cos\phi\partial_\theta-\cot\theta\sin\phi\partial_\phi,\nonumber\\
\xi_{(3)}&=&\partial_\phi,\nonumber\\
\xi_{(4)}&=&\psi\frac{T_0}{T'_0}\partial_r,\nonumber\\
\xi_{(5)}&=&k_1(\cos t\partial_t+\frac{2T_0}{T'_0}\sin t
\partial_r)\sin\theta\sin\phi\nonumber\\
&-&(\cos\theta\sin\phi\partial_\theta
+\csc\theta\cos\phi\partial_\phi)\cos t,\nonumber\\
\xi_{(6)}&=&k_1(\sin t\partial_t-\frac{2T_0}{T'_0}\cos t
\partial_r)\sin\theta\sin\phi\nonumber\\
&-&(\cos\theta\sin\phi\partial_\theta
+\csc\theta\cos\phi\partial_\phi)\sin t,\nonumber\\
\xi_{(7)}&=&k_1(\cos t\partial_t+\frac{2T_0}{T'_0}\sin t
\partial_r)\sin\theta\cos\phi\nonumber\\
&-&(\cos\theta\cos\phi\partial_\theta
+\csc\theta\sin\phi\partial_\phi)\cos t,\nonumber\\
\xi_{(8)}&=&k_1(\sin t\partial_t-\frac{2T_0}{T'_0}\cos t
\partial_r)\sin\theta\cos\phi\nonumber\\
&-&(\cos\theta\cos\phi\partial_\theta
+\csc\theta\sin\phi\partial_\phi)\sin t,\nonumber\\
\xi_{(9)}&=&k_1(\cos t\partial_t+\frac{2T_0}{T'_0}\sin t
\partial_r)\cos\theta+\cos t\sin\theta\partial_\theta,\nonumber\\
\xi_{(10)}&=&k_1(\sin t\partial_t-\frac{2T_0}{T'_0}\cos t
\partial_r)\cos\theta+\sin t\sin\theta\partial_\theta.
\end{eqnarray}
Thus we obtain eleven independent MI vectors in this degenerate
case in which four are the usual isometries of the spherically
symmetric spacetime.

The case 2$a_1b_2(c_2)$ yields the following five independent MIC
\begin{eqnarray}
\xi_{(0)}&=&(\frac{\psi(k_2-1)}{2k_2}t+1)\partial_t,\nonumber\\
\xi_{(1)}&=&\sin\phi\partial_\theta+\cot\theta\cos\phi\partial_\phi,\nonumber\\
\xi_{(2)}&=&\cos\phi\partial_\theta-\cot\theta\sin\phi\partial_\phi,\nonumber\\
\xi_{(3)}&=&\partial_\phi,\nonumber\\
\xi_{(4)}&=&\psi\frac{T_0}{T'_0}\partial_r.
\end{eqnarray}
where $T'_2\neq0,~T_2=k_1T_0^{k_2},~k_1\neq0,~k_2\neq0,1$. Thus we
obtain finite dimensional MICs in another degenerate case.\\
\par\noindent
{\bf Subcase 2$(a_2)$:} When we solve MIC Eqs.(8), (9), (14) and
(20) together with the constraint of this case, we have
\begin{equation}
T_0(\xi^0_{,23}-\cot\theta\xi^0_{,3})=0,
\end{equation}
From this equation, we further have the following two
possibilities:
\begin{eqnarray*}
(b_1)\quad T_0=0,\quad \xi^0_{,23}-\xi^0_{,3}\cot\theta\neq0,\nonumber\\
\quad (b_2)\quad T_0\neq0,\quad
\xi^0_{,23}-\xi^0_{,3}\cot\theta=0.
\end{eqnarray*}
For the case 2$a_2(b_1)$, when we solve MIC equations
simultaneously for these constraints, we obtain the following
solution
\begin{eqnarray}
\xi_{(0)}&=&\xi^0(x^a)\partial_t,\nonumber\\
\xi_{(1)}&=&\sin\phi\partial_\theta+\cot\theta\cos\phi\partial_\phi,\nonumber\\
\xi_{(2)}&=&\cos\phi\partial_\theta-\cot\theta\sin\phi\partial_\phi,\nonumber\\
\xi_{(3)}&=&\partial_\phi,\nonumber\\
\xi_{(4)}&=&\psi\frac{T_2}{T'_2},\nonumber\\
\xi_{(5)}&=&\frac{2\sqrt{T_2}}{T'_2}\sin\theta\sin\phi\partial_r
+\frac{1}{\sqrt{T_2}}\cos\theta\sin\phi\partial_\theta
+\frac{1}{\sqrt{T_2}}\csc\theta\cos\phi\partial_\phi,\nonumber\\
\xi_{(6)}&=&\frac{2\sqrt{T_2}}{T'_2}\sin\theta\cos\phi\partial_r
+\frac{1}{\sqrt{T_2}}\cos\theta\cos\phi\partial_\theta
+\frac{1}{\sqrt{T_2}}\csc\theta\sin\phi\partial_\phi,\nonumber\\
\xi_{(7)}&=&\frac{2\sqrt{T_2}}{T'_2}\cos\theta\partial_r
-\frac{1}{\sqrt{T_2}}\sin\theta\partial_\theta,
\end{eqnarray}
where $4T_1T_2=T'^2_2$. This gives infinite dimensional MICs.\\
\par\noindent
The case 2$a_2(b_2)$, i.e., $T_0\neq0,\quad
\xi^0_{,23}-\xi^0_{,3}\cot\theta=0$ turns out to be
non-degenerate. When we use these constraints in the MIC
Eqs.(6)-(15) and solve them simultaneously, after some tedious
algebra, we arrive at the following solution
\begin{equation}
\xi^0=-\frac{T_2}{T_0}[(\dot{A_1}\sin\phi
-\dot{A_2}\cos\phi)\sin\theta+\dot{A_3}\cos\theta]+A_4,
\end{equation}
\begin{equation}
\xi^1=-\frac{T_2}{T_1}[(A'_1\sin\phi-A'_2\cos\phi)\sin\theta
+A'_3\cos\theta]+A_5,
\end{equation}
\begin{equation}
\xi^2=(A_1\sin\phi-A_2\cos\phi)\cos\theta+A_3\sin\theta
+c_1\sin\phi-c_2\cos\phi,
\end{equation}
\begin{equation}
\xi^3=(A_1\cos\phi+A_2\sin\phi){\rm cosec}\theta
+(c_1\cos\phi+c_2\sin\phi)\cot\theta+c_0,
\end{equation}
where $c_0, c_1$ and $c_2$ are arbitrary constants and
$A_\mu=A_\mu(t,r),~\mu=1,2,3,4,5$ are arbitrary functions. Here
dot and prime indicate the differentiation with respect to time
and $r$ coordinate respectively. When we substitute these values
of $\xi^a$ in MIC Eqs.(6)-(15), we obtain the following
differential constraints on $A_\nu$ with $c_0=0$. These $\xi^a$
are satisfied subject to the following differential constraints on
$A_\mu$
\begin{eqnarray}
2T_1\ddot{A}_i+T_{0,1}A'_i=0,~~i = 1,2,3,\\
2T_0 \dot{A_4}+T_{0,1}A_5 =  \psi T_0,\\
2T_2 \dot{A'_i} + T_{0} \left({T_{2} \over T_{0} } \right)'
\dot{A_i}= 0,\\
T_{0} A'_4 + T_1\dot{A_5} = 0,\\
\left\{T_{1,1} {T_2 \over T_1} + 2 T_1 \left({ T_2
\over T_1} \right)' \right\} A'_i + 2 T_2 A_i^{''} = 0,\\
T_{1,1} A_5 + 2 T_1 A'_5 = \psi T_1,\\
T_{2,1} A'_i + 2 T_1 A_i = 0,~~c_0 = 0,\\
T_{2,1} A_5 =\psi T_2.
\end{eqnarray}
Now the problem of working out MICs for all possibilities of $
A_i, A_4, A_5 $ is reduced to solving the set of Eqs.(28)-(31)
subject to the above constraints. We start the classification of
MICs by considering the constraint Eq.(39). It follows that $A_5$
is a function of coordinate $r$ only and
\begin{equation}
\psi=\frac{T'_2A_5(r)}{T_2}.
\end{equation}
When we use this value in Eqs.(33), (35) and (37) and solve
simultaneously, we obtain
\begin{equation}
A_4(t)=\frac{k}{2}[\frac{T_0T'_2-T'_0T_2}{T_0\sqrt{T_1T_2}}]t+c_0,
\end{equation}
where $k$ is an integration constant. When we differentiate this
equation with respect to time, we have
\begin{equation}
T_0T'_2-T'_0T_2=mT_0\sqrt{T_1T_2},
\end{equation}
where $m$ is an arbitrary constant which can be zero or non-zero.
\begin{eqnarray*}
(c_1)\quad m=0,\quad (c_2)\quad m\neq0.
\end{eqnarray*}
The case 2$a_2b_2(c_1)$ gives
\begin{equation}
A_i=(\frac{a_it^2}{2}+b_it)-a_i\int{\frac{2T_1}{T'_0}dr}+c_i.
\end{equation}
Now if we use Eqs.(38), (40) and (41), we finally obtain the
following MIC
\begin{eqnarray}
\xi_{(0)}&=&\partial_t,\nonumber\\
\xi_{(1)}&=&\sin\phi\partial_\theta+\cot\theta\cos\phi\partial_\phi,\nonumber\\
\xi_{(2)}&=&\cos\phi\partial_\theta-\cot\theta\sin\phi\partial_\phi,\nonumber\\
\xi_{(3)}&=&\partial_\phi,\nonumber\\
\xi_{(4)}&=&\psi\frac{T_0}{T'_0}
\end{eqnarray}
which gives five MI vectors.

For the case 2$a_2b_2(c_2)$ when $m\neq0$, we use Eqs.(32), (41)
and (42) to obtain
\begin{equation}
\frac{\ddot{B}_i(t)}{B_i(t)}=-\frac{T'_0(\sqrt{\frac{T_0}{T_2}})'}
{2T_2\sqrt{\frac{T_0}{T_2}}}=\alpha_i,
\end{equation}
where $\alpha_i$ is a separation constant which yields the
following three different possibilities:
\begin{eqnarray*}
(d_1)\quad \alpha_i>0,\quad (d_2)\quad \alpha_i=0,\quad (d_3)\quad
\alpha_i<0.
\end{eqnarray*}
In the case 2$a_2b_2c_2(d_1)$, i.e., $\alpha_i>0$, when we solve
Eqs.(32), (34), (36) and (38), we obtain
\begin{equation}
[\frac{T'_1}{2T_1}+\frac{T_1}{T_2}(\frac{T_2}{T_1})']
(\sqrt{\frac{T_0}{T_2}})'+(\sqrt{\frac{T_0}{T_2}})''
[(a_i\sinh\sqrt{\alpha_i}t+b_i\cosh\sqrt{\alpha_i}t\frac{1}
{\sqrt{\alpha_i}}]=0
\end{equation}
It follows from the above equation that either
\begin{eqnarray*}
(e_1)\quad [\frac{T'_1}{2T_1}+\frac{T_1}{T_2}(\frac{T_2}{T_1})']
(\sqrt{\frac{T_0}{T_2}})'+(\sqrt{\frac{T_0}{T_2}})''=0;\quad
a_i\neq0\neq b_i,\\
or\quad(e_2)\quad
[\frac{T'_1}{2T_1}+\frac{T_1}{T_2}(\frac{T_2}{T_1})']
(\sqrt{\frac{T_0}{T_2}})'+(\sqrt{\frac{T_0}{T_2}})''\neq0;\quad
a_i=0=b_i.
\end{eqnarray*}
However, both the above cases 2$a_2b_2c_2d_1(e_1)$ and
2$a_2b_2c_2d_1(e_2)$ give contradiction and hence can be excluded.

In the case 2$a_2b_2c_2(d_2)$, i.e., $\alpha_i=0$ when we use
constraint equations, we obtain the following MIC
\begin{eqnarray}
\xi_{(0)}&=&\partial_t,\nonumber\\
\xi_{(1)}&=&\sin\phi\partial_\theta+\cot\theta\cos\phi\partial_\phi,\nonumber\\
\xi_{(2)}&=&\cos\phi\partial_\theta-\cot\theta\sin\phi\partial_\phi,\nonumber\\
\xi_{(3)}&=&\partial_\phi,\nonumber\\
\xi_{(4)}&=&\sqrt{T_2}\sin\theta\sin\phi\partial_t
-\frac{2\sqrt{T_2}}{T'_2}t\sin\theta\sin\phi\partial_r
-\frac{1}{\sqrt{T_2}}t\cos\theta\sin\phi\partial_\theta,\nonumber\\
\xi_{(5)}&=&\sqrt{T_2}\sin\theta\cos\phi\partial_t
-\frac{2\sqrt{T_2}}{T'_2}t\sin\theta\cos\phi\partial_r
-\frac{1}{\sqrt{T_2}}t\cos\theta\cos\phi\partial_\theta,\nonumber\\
\xi_{(6)}&=&\sqrt{T_2}\cos\theta\partial_t
-\frac{2\sqrt{T_2}}{T'_2}t\cos\theta\partial_r
+\frac{1}{\sqrt{T_2}}t\sin\theta\partial_\theta,
+\frac{1}{\sqrt{T_2}}t\csc\theta\cos\phi\partial_\phi,\nonumber\\
\xi_{(7)}&=&\frac{2\sqrt{T_2}}{T'_2}\sin\theta\sin\phi\partial_r
+\frac{1}{\sqrt{T_2}}\cos\theta\sin\phi\partial_\theta
+\frac{1}{\sqrt{T_2}}\csc\theta\cos\phi\partial_\phi,\nonumber\\
\xi_{(8)}&=&\frac{2\sqrt{T_2}}{T'_2}\sin\theta\cos\phi\partial_r
+\frac{1}{\sqrt{T_2}}\cos\theta\cos\phi\partial_\theta
+\frac{1}{\sqrt{T_2}}\csc\theta\sin\phi\partial_\phi,\nonumber\\
\xi_{(9)}&=&\frac{2\sqrt{T_2}}{T'_2}\cos\theta\partial_r
-\frac{1}{\sqrt{T_2}}\sin\theta\partial_\theta,\nonumber\\
\xi_{(10)}&=&\psi\frac{T_2}{T'_2}\partial_r
\end{eqnarray}
which gives eleven MI vectors.

The case 2$a_2b_2c_2(d_3)$, i.e., $\alpha_i<0$ gives similar
results as the case 2$a_2b_2c_2(d_1)$.

\section{Conclusion}

We know that when the spherically symmetric static spacetimes are
classified according to their MCs [11], we obtain either {\it
four}, {\it six}, {\it seven} or {\it ten} independent MCs for the
non-degenerate energy-momentum tensor. For the degenerate case, we
obtain infinite dimensional MCs except the two cases where the
energy-momentum tensor is degenerate but the group of matter
collineations is finite-dimensional, i.e. {\it four} or {\it ten}.

In this paper, we have classified spherically symmetric spcetimes
according to their MICs. We see that for the degenerate
energy-momentum tensor, most of the cases yield infinite
dimensional MICs. The worth mentioning cases are those where we
have got finite number of MICs even when the energy-momentum
tensor is zero. We obtain two such different cases having either
eleven or five independent MICs. In the non-degenerate case, we
obtain either five or eleven independent MICs. These contain the
usual four isometries of the spherically symmetry and the rest are
the proper (non-trivial) MICs. It can easily be seen from these
cases that if we take the inheriting factor $\psi=0$, MICs reduce
to MCs. These MCs exactly coincide with those already found in the
literature [11]. It is mentioned here that MICs coincide with RICs
[18] but the constraint equations are completely different. We
conclude that the idea of matter inheritance collineations turns
out to be the same as homotheties and conformal Killing vectors
for the metric tensor. The results are summarized in the form of
tables given below.

\newpage
{\bf {\small Table 1.}} {\small MICs for Spherically Symmetric
Static Spacetimes}

\vspace{0.1cm}

\begin{center}
\begin{tabular}{|l|l|l|}
\hline {\bf Cases} & {\bf MICs} & {\bf Constraints}
\\ \hline 1$a_1(b_1)$ & Infinte No. of MICs & $
\begin{array}{c}
T_2=0,~\xi^2_{,3}+\xi^3_{,2}\sin^2\theta\neq0,~T_1=0,\\
\xi^1_{,2}\neq0\neq\xi^1_{,3},~T_0=0,\\
\xi^0_{,2}\neq0\neq\xi^0_{,3}\neq\xi^0_{,1}
\end{array}
$\\ \hline 1$a_1(b_2)$ & Contradiction & $
\begin{array}{c}
T_2=0,~\xi^2_{,3}+\xi^3_{,2}\sin^2\theta\neq0,~T_1=0,\\
\xi^1_{,2}\neq0\neq\xi^1_{,3},~T_0\neq0,\\
\xi^0_{,2}=0=\xi^0_{,3}=\xi^0_{,1}
\end{array}
$\\ \hline 1$a_2(b_1)$ & Infinte No. of MICs & $
\begin{array}{c}
T_2=0,~\xi^2_{,3}+\xi^3_{,2}\sin^2\theta\neq0,~T_1\neq0,\\
\xi^1_{,2}=0=\xi^1_{,3},~T_0=0,~ \xi^0_{,2}\neq0\neq\xi^0_{,3}
\end{array}
$\\ \hline 1$a_2b_2(c_1)$ & Infinte No. of MICs & $
\begin{array}{c}
T_2=0,~\xi^2_{,3}+\xi^3_{,2}\sin^2\theta\neq0,~T_1\neq0,\\
\xi^1_{,2}=0=\xi^1_{,3},~T_0\neq0,~
\xi^0_{,2}=0=\xi^0_{,3},\\
\frac{T_0}{T_1}(\frac{T'_0}{2T_0\sqrt{T_1}})'=k,~k>0
\end{array}
$\\ \hline 1$a_2b_2c_2(d_1)$ & Contradiction & $
\begin{array}{c}
T_2=0,~\xi^2_{,3}+\xi^3_{,2}\sin^2\theta\neq0,~T_1\neq0,\\
\xi^1_{,2}=0=\xi^1_{,3},~T_0\neq0,~ \xi^0_{,2}=0=\xi^0_{,3},\\
k=0,\frac{T'_0}{2T_0\sqrt{T_1}}=m,~m>0
\end{array}
$\\ \hline 1$a_2b_2c_2(d_2)$ & Infinite No. of MICs& $
\begin{array}{c}
T_2=0,~\xi^2_{,3}+\xi^3_{,2}\sin^2\theta\neq0,~T_1\neq0,\\
\xi^1_{,2}=0=\xi^1_{,3},~T_0\neq0,~
\xi^0_{,2}=0=\xi^0_{,3},\\
k=0,~m=0
\end{array}
$\\ \hline 1$a_2b_2c_2(d_3)$ & Contradiction & $
\begin{array}{c}
T_2=0,~\xi^2_{,3}+\xi^3_{,2}\sin^2\theta\neq0,~T_1\neq0,\\
\xi^1_{,2}=0=\xi^1_{,3},~T_0\neq0,~
\xi^0_{,2}=0=\xi^0_{,3},\\
k=0,~m<0
\end{array}
$\\ \hline 2$a_1(b_1)$ & Infinite No. of MICs & $
\begin{array}{c}
T_2\neq0,~\xi^2_{,3}+\xi^3_{,2}\sin^2\theta=0,~T_1=0,\\
\xi^1_{,23}-\xi^1_{,3}\cot\theta\neq0,~T_0=0,~\xi^0_{,1}\neq0,
\end{array}
$\\ \hline 2$a_1b_2(c_1)$ & $11$ & $
\begin{array}{c}
T_2\neq0,~\xi^2_{,3}+\xi^3_{,2}\sin^2\theta=0,~T_1=0,\\
\xi^1_{,23}-\xi^1_{,3}\cot\theta\neq0,~T_0\neq0,~\xi^0_{,1}=0,\\
(\frac{T_0}{T_2})'=0,~\xi^0_{,2}\neq0
\end{array}
$\\ \hline 2$a_1b_2(c_2)$ & $5$ & $
\begin{array}{c}
T_2\neq0,~\xi^2_{,3}+\xi^3_{,2}\sin^2\theta=0,~T_1=0,\\
\xi^1_{,23}-\xi^1_{,3}\cot\theta\neq0,~T_0\neq0,~\xi^0_{,1}=0,\\
(\frac{T_0}{T_2})'\neq0,~\xi^0_{,2}=0
\end{array}
$\\ \hline 2$a_2(b_1)$ & Infinte No. of MICs & $
\begin{array}{c}
T_2\neq0,~\xi^2_{,3}+\xi^3_{,2}\sin^2\theta=0,\\
T_1\neq0,~\xi^1_{,23}-\xi^1_{,3}\cot\theta=0,
\\T_0=0,~\xi^0_{,23}-\xi^0_{,3}\cot\theta\neq0
\end{array}
$\\ \hline
\end{tabular}
\end{center}
\begin{center}
\begin{tabular}{|l|l|l|}
\hline {\bf Cases} & {\bf MICs} & {\bf Constraints}
\\ \hline 2$a_2b_2(c_1)$ & $5$ & $
\begin{array}{c}
T_2\neq0,~\xi^2_{,3}+\xi^3_{,2}\sin^2\theta=0,\\
T_1\neq0,~\xi^1_{,23}-\xi^1_{,3}\cot\theta=0,\\
T_0\neq0,~\xi^0_{,23}-\xi^0_{,3}\cot\theta=0,\\
T_0T'_2-T'_0T_2=mT_0\sqrt{T_1T_2},~m=0
\end{array}
$\\ \hline 2$a_2b_2c_2d_1(e_1)$ & Contradiction & $
\begin{array}{c}
T_2\neq0,~\xi^2_{,3}+\xi^3_{,2}\sin^2\theta=0,\\
T_1\neq0,~\xi^1_{,23}-\xi^1_{,3}\cot\theta=0,\\
T_0\neq0,~\xi^0_{,23}-\xi^0_{,3}\cot\theta=0,\\
m\neq0,~-\frac{T'_0(\sqrt{\frac{T_0}{T_2}})'}{2T_2\sqrt{T_0}{T_2}}
=\alpha_i,~\alpha_i>0,\\
(\frac{T'_1}{2T_1}+\frac{T_1}{T_2}(\frac{T_2}{T_1})')
(\sqrt{\frac{T_0}{T_2}})'\\
+(\sqrt{\frac{T_0}{T_2}})''=0,~a_i\neq0\neq b_i
\end{array}
$\\ \hline 2$a_2b_2c_2d_1(e_2)$ & Contradiction & $
\begin{array}{c}
T_2\neq0,~\xi^2_{,3}+\xi^3_{,2}\sin^2\theta=0,\\
T_1\neq0,~\xi^1_{,23}-\xi^1_{,3}\cot\theta=0,\\
T_0\neq0,~\xi^0_{,23}-\xi^0_{,3}\cot\theta=0,\\
m\neq0,~\alpha_i>0,\\
(\frac{T'_1}{2T_1}+\frac{T_1}{T_2}(\frac{T_2}{T_1})')
(\sqrt{\frac{T_0}{T_2}})'\\
+(\sqrt{\frac{T_0}{T_2}})''\neq0,~a_i=0=b_i
\end{array}
$\\ \hline 2$a_2b_2c_2(d_2)$ & $11$ & $
\begin{array}{c}
T_2\neq0,~\xi^2_{,3}+\xi^3_{,2}\sin^2\theta=0,\\
T_1\neq0,~\xi^1_{,23}-\xi^1_{,3}\cot\theta=0,\\
T_0\neq0,~\xi^0_{,23}-\xi^0_{,3}\cot\theta=0,\\
m\neq0,~\alpha_i=0
\end{array}
$\\ \hline 2$a_2b_2c_2(d_3)$ & 2$a_2b_2c_2(d_1)$ & $
\begin{array}{c}
T_2\neq0,~\xi^2_{,3}+\xi^3_{,2}\sin^2\theta=0,\\
T_1\neq0,~\xi^1_{,23}-\xi^1_{,3}\cot\theta=0,\\
T_0\neq0,~\xi^0_{,23}-\xi^0_{,3}\cot\theta=0,\\
m\neq0,~\alpha_i<0
\end{array}
$\\ \hline
\end{tabular}
\end{center}
It follows from these tables that each case has different
constraints on the energy-momentum tensor. It would be interesting
to solve these constraints to ascertain the non-trivial MIC.

\newpage

{\bf \large References}

\begin{description}

\item{[1]} Duggal, K.L.: J. Math. Phys., {\bf 33}(1992)2989.

\item{[2]} Duggal, K.L.: Acta Applicanade Mathematica, {\bf 31}(1993)225.

\item{[3]} Hall, G.S., Roy, I. and Vaz, L.R.: Gen. Rel and Grav.
{\bf 28}(1996)299.

\item{[4]} Camc{\i}, U. and Barnes, A.: Class. Quant. Grav. {\bf
19}(2002)393.

\item{[5]} Carot, J. and da Costa, J.: {\it Procs. of the 6th
Canadian Conf. on General Relativity and Relativistic
Astrophysics}, Fields Inst. Commun. 15, Amer. Math. Soc. WC
Providence, RI(1997)179.

\item{[6]} Carot, J., da Costa, J. and Vaz, E.G.L.R.: J. Math.
Phys. {\bf 35}(1994)4832.

\item{[7]} Sharif, M.: Nuovo Cimento {\bf B116}(2001)673;\\
Astrophys. Space Sci. {\bf 278}(2001)447.

\item{[8]} Camc{\i}, U. and Sharif, M.: Gen Rel. and Grav. {\bf
35}(2003)97.

\item{[9]} Camc{\i}, U. and Sharif, M.: Class. Quant. Grav.
{\bf 20}(2003)2169-2179.

\item{[10]} Tsamparlis, M. and Apostolopoulos, P.S.: Gen. Rel.
and Grav. {\bf 36}(2004)47.

\item{[11]} Sharif, M. and Sehar Aziz: Gen Rel. and Grav. {\bf
35}(2003)1091;\\
Sharif, M.: J. Math. Phys. {\bf 44}(2003)5142.

\item{[12]} Sharif, M.: J. Math. Phys. {\bf 45}(2004)1518;

\item{[13]} Sharif, M.: J. Math. Phys. 45(2004)1532.

\item{[14]} Katzin, G.H., Levine J. and Davis, W.R.: J. Math. Phys.
{\bf 10}(1969)617.

\item{[15]} Petrov, A.Z.: {\it Einstein Spaces} (Pergamon, Oxford
University Press, 1969).

\item{[16]} Stephani, H., Kramer, D., MacCallum, M.A.H.,
Hoenselaers, C. and Hearlt, E.: {\it Exact Solutions of Einstein's
Field Equations} (Cambridge University Press, 2003).

\item{[17]} Rcheulishrili, G.: J. Math. Phys. {\bf 33}(1992)1103.

\item{[18]} Bokhari, A.H., Kashif, A.R. and Kara A.H.: Nuovo Cimento
{\bf 118B}(2003)803.

\end{description}

\end{document}